\documentclass[pra,twocolumn,showpacs,nobibnotes,floatfix,superscriptaddress]{revtex4}

\usepackage{latexsym,amsmath,amssymb,amsfonts,mathbbol,graphicx,color}
\usepackage{dcolumn}
\usepackage{bm}

\begin{document}

\title{The Fidelity of an Encoded $[7,1,3]$ Logical Zero} 
\author{Yaakov S. Weinstein}
\affiliation{Quantum Information Science Group, {\sc Mitre},
260 Industrial Way West, Eatontown, NJ 07724, USA}

\begin{abstract}
I calculate the fidelity of a $[7,1,3]$ CSS quantum error correction code 
logical zero state constructed in a non-equiprobable Pauli operator error 
environment for two methods of encoding. The first method is to apply fault tolerant
error correction to an arbitrary state of 7 qubits utilizing Shor states for syndrome 
measurement. The Shor states are themselves constructed in the 
non-equiprobable Pauli operator error environment and their fidelity 
depends on the number of verifications done to ensure multiple errors 
will not propagate into the encoded quantum information. Surprisingly, performing 
these verifications may lower the fidelity of the constructed Shor states. 
The second encoding method is to simply implement the $[7,1,3]$ encoding 
gate sequence also in the non-equiprobable Pauli operator error environment. 
Perfect error correction is applied after both methods to determine the 
correctability of the implemented errors. I find that which method attains
a higher fidelity depends on the which of Pauli operator errors is dominant. 
Nevertheless, perfect error correction supresses errors to at least first order
for both methods. 
\end{abstract}


\maketitle

\section{Introduction}
Quantum error correction (QEC) will be vital for the proper working of any hoped-for 
quantum computer \cite{book}. QEC codes must be more elaborate than classical codes due to the need 
to correct both bit-flip and phase-flip errors \cite{ShorQEC}. While such codes exist \cite{CSS} 
their existence alone is not sufficient to guarantee the theoretical possibility of constructing 
a working quantum computer. Rather a framework is required to ensure accurate manipulation and 
recovery of quantum information despite the inevitable presence of errors. This framework has 
been painstakingly built over the last 15 years and is called quantum fault tolerance 
\cite{G,ShorQFT,Preskill,AGP}. 

Here, we carefully examine a basic component of this fault tolerant framework for the $[7,1,3]$ QEC 
code: constructing an encoded logical zero. The $[7,1,3]$ code, or Steane code \cite{Steane}, fully 
protects one qubit of quantum 
information from bit-flip and phase-flip errors by embedding it into a system of seven qubits. Error 
correction is implemented by performing controlled-NOT (CNOT) gates between the 7 data qubits and 
ancilla qubits which are measured to determine the error syndrome. Based on the outcome of the syndrome 
measurements the appropriate recovery operation is applied. To assure fault tolerance, gates must be 
applied in such a way as to ensure errors will not propagate in an uncontrolled fashion during algorithm 
implementation. In addition, the code can be concatenated \cite{conc} to decrease the tolerable error
probability threshold. 

A method of encoding the state $|0\rangle$ into the Steane code in a fault tolerant manner is 
to apply fault tolerant error correction to any initial state of 7 qubits. This requires construction 
of proper ancilla syndrome qubits such that each ancilla interacts with no more than one data qubit, 
thereby stemming uncontrolled propagation of possible ancilla qubit errors. For the Steane code 
appropriate ancilla qubits are four qubit Shor states \cite{ShorQFT}. Shor states are simply GHZ states 
with Hadamard gates applied to each qubit. A second encoding method is to utilize Steane's encoding 
gate sequence \cite{Steane}. This method is not fault tolerant because an error in the encoding 
sequence can propagate to multiple qubits. 

In this paper I compare the accuracy with which the two methods encode logical zero states 
using a reasonable, but general, error model: a non-equiprobable Pauli 
operator error environment \cite{QCC}. As in \cite{AP}, this model is a stochastic version of 
a biased noise model that can be formulated in terms of Hamiltonians coupling the system to an 
environment. In the model used here, however, the probabilities with which the different error 
types will take place is left arbitrary: the environment causes qubits to undergo a $\sigma_x^j$ error 
with probability $p_x$, a $\sigma_y^j$ error with probability $p_y$, and a $\sigma_z^j$ error 
with probability $p_z$, where $\sigma_i^j$, $i = x,y,z$ are the Pauli spin 
operators on qubit $j$. I assume that only qubits taking part in a gate will undergo error and 
the error is modeled to occur after (perfect) gate implementation. Qubits not involved in a gate are 
assumed to be perfectly stored. While this represents an idealization, it is justifiable as all 
accuracy measures are calculated only to second order in the error probabilities $p_i$. I also 
make the simplification that possible errors in preparation and measurement can be assumed to 
have occurred in subsequent or immediately preceding gates.

After (noisy) construction of the logical zero state I apply noiseless quantum error correction.
The accuracy of the error corrected state tells us whether or not the errors that had occurred 
during the encoding are, in principle, correctable. An encoding method that produces
uncorrectable errors with probabilities first order in the $p_i$
would presumably be unusable for practical, fault tolerant, quantum computation. 

A previous comparison of these two construction methods using the less general equiprobable error 
model was done in Ref.~\cite{SDF}. There are other significant distinctions between that work and 
what is considered here. In the current work Shor states are explicitly constructed and verified
and their accuracy is determined. These (necessarily) noisy Shor states are then used for syndrome 
measurement in the fault tolerant encoding method. In addition, the fidelity used in this work
as an accuracy measure is the fidelity with respect only to the desired state (logical zero). Due 
to these differences, the results I present here do not reduce to those of Ref.~\cite{SDF}.

\section{Shor State Construction}

\begin{figure}[t]
\includegraphics[width=8cm]{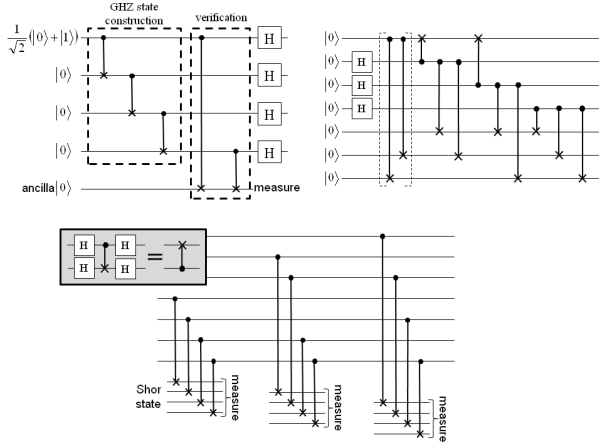}
\caption{Top left: construction of a 4 qubit Shor state. CNOT gates are represented by dots on the control qubit 
and $\times$ on the target qubit connected by a line. $H$ represents a Hadamard gate. The procedure
entails constructing a GHZ state which is verified using ancilla qubits. Hadamard gates are 
applied to each qubit to complete Shor state construction.
Bottom: syndrome measurements for the $[7,1,3]$ code. For fault tolerant syndrome measurement each ancilla 
qubit must interact with only one data qubit. The error syndrome is determined from the parity of the 
measurement outcomes of the Shor state. To achieve fault tolerance each of the syndrome measurements is repeated twice. 
Box: a useful equality which allows us to avoid implementing Hadamard gates by reversing 
the control and target of CNOT gates. In our context, the CNOTs are reversed such that the ancilla 
qubits become the control and the data qubits become the target, as explained in the text. 
Top right: gate sequence for encoding an arbitrary state, stored on the first qubit, into the 
Steane code shown here for the initial state $|0\rangle$. For this initial state the first two 
CNOTs (shown in brackets) need not be implemented. }
\label{ShorState}
\end{figure}

To implement fault tolerant encoding of a logical zero state, it is first necessary to construct 
Shor states for the purpose of syndrome measurement. Perfect Shor states (without the final 
Hadamard gates as explained below) are given by 
$|\psi_{Shor}\rangle = \frac{1}{\sqrt{2}}(|0000\rangle + |1111\rangle)$. For our simulations 
we must determine what the Shor state will look like when constructed in the non-equiprobable error 
environment. These states can then be used for our simulations of syndrome measurements of 
the 7 data qubits which are to make up the logical zero. To construct the Shor state we start 
with four qubits assumed to be perfectly initialized in the state $\rho_i = |\psi_i\rangle\langle\psi_i|$, where 
$|\psi_i\rangle = \frac{1}{\sqrt{2}}(|0\rangle+|1\rangle)|000\rangle$. 
As each gate is applied, the non-equiprobable error environment causes the qubits taking 
part in the gate to probabilistically undergo errors as described above. Any attempted 
performance of a unitary CNOT gate with control qubit $j$ and target qubit $k$, $C_jNOT_k$, 
on any state $\rho$ actually implements: 
\begin{equation}
\sum_{a,b}^{0,x,y,z}p_ap_b\sigma_a^j\sigma_b^kC_jNOT_k\rho C_jNOT_k^{\dag}\sigma_a^j\sigma_b^k,
\label{cnot}
\end{equation}
where $\sigma_0^j$ is the identity matrix, $p_0 = 1-\sum_{\ell=x,y,z}p_\ell$, and 
the terms $A_{a,b}^{j,k} = \sqrt{p_ap_b}\sigma_a^j\sigma_b^kC_jNOT_k$ can be regarded as Kraus operators. 
Shor state construction requires three CNOT gates, shown in Fig.~\ref{ShorState}, and thus the final state 
is given by
\begin{equation}  
\rho_{Shor-err} = \sum_{a,b,c,d,e,f}^{0,x,y,z}A_{e,f}^{3,4}A_{c,d}^{2,3}A_{a,b}^{1,2}\rho_i(A_{a,b}^{1,2})^\dag(A_{c,d}^{2,3})^\dag(A_{e,f}^{3,4})^\dag.
\end{equation}

To quantify the accuracy of the constructed Shor state we use the fidelity given 
by $F = \langle\psi_{Shor}|\rho_{Shor-err}|\psi_{Shor}\rangle$. 
For the non-equiprobable error environment we find that the fidelity of the constructed
Shor state is:
\begin{eqnarray}
F_{noAnc} &=& 1-6p_x-6p_y-6p_z+16p_x^2+30p_xp_y \nonumber\\
	    &+& 30p_xp_z+16p_y^2+30p_yp_z+30p_z^2.
\end{eqnarray}
Since we have ignored higher order terms, this expression is accurate only for 
small values of $p_i$.

Were we to simply apply the above described gate sequence we would have not built the Shor states 
in a fault tolerant fashion. This is because multiple errors in Shor state construction 
can propagate into the data qubits when the Shor states are used for syndrome measurement. 
We need to test the Shor states to ensure that multiple 
errors have not taken place. This is done utilizing an ancilla qubit, initially in the 
state $|0\rangle$, adjoined to the Shor state to measure the parity of random pairs of 
qubits \cite{ShorQFT}. Should the test fail (the ancilla qubit measurement yields a 
$|1\rangle$), the Shor state is immediately discarded. Of course, the CNOT gates used 
in this parity measure are themselves performed in the non-equiprobable
error environment and their dynamics follows that of Eq.~\ref{cnot}. 

\begin{figure}[t]
\includegraphics[width=8cm]{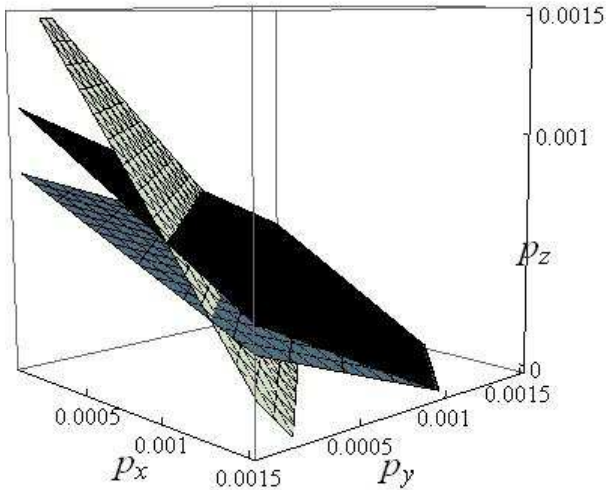}
\caption{Contour plots comparing the fidelity of three methods of constructing Shor states, 
no ancilla-based verifications (white), one verification between qubits 1 and 4 
(black), and two verifications, the first between qubits 1 and 4 and the second 
between qubits 1 and 2 (gray). The contours show where in the error space ($p_x, p_y$, $p_z$)
the fidelity of the constructed Shor state is equal to .99. Note that the origin (where the 
three probabilities equal zero) is at the lower back left corner. The once verified Shor 
state achieves a fidelity of .99 at higher error probabilities than the twice verified 
Shor state except in the case of very low $p_z$. For higher values of $p_z$ the 
unverified state achieves a fidelity of .99 at the highest error probabilities. }
\label{fids1}
\end{figure}

I utililze an initial ancilla qubit to measure the parity of qubits 1 and 4. 
After application of the CNOT gates the ancilla will measure $|0\rangle$ with probability (to second
order) of $1-6p_x-6p_y+30p_x^2+30p_y^2+60p_xp_y$. The resulting Shor state fidelity is now
\begin{eqnarray}
F_{1Anc} &=& 1-4p_x-4p_y-9p_z-7p_x^2-17p_xp_y \nonumber\\
	    &+& 27p_xp_z-8p_y^2+27p_yp_z+72p_z^2.
\end{eqnarray}
While the verification should work to ensure that errors cannot propagate into the data
qubits, it increases the fidelity of the constructed Shor state only for certain values of 
the $p_i$. However, when $\sigma_z$ is the dominant error operator the fidelity
of the Shor state is higher when no verification is done as shown in Fig.~\ref{fids1}. Thus, 
one must question whether implementing the verification is the correct approach for these 
error values.

Applying additional verification steps using additional ancilla should further ensure the lack 
of errors in the constructed Shor states. A second ancilla 
can recheck the parity of the qubits checked with the previous ancilla, or check the parity of 
additional Shor state qubits. I find that the Shor state fidelity is highest when checking 
qubits in which at least one is the same as that checked by the first ancilla. For example, 
a second ancilla can be used to check the parity between qubits 1 and 2. In this case, the 
ancilla will measure $|0\rangle$ with probability (up to second order) 
$1-4p_x-4p_y+4p_x^2+4p_y^2+8p_xp_y$. The resulting density matrix has a fidelity 
(to second order) of: 
\begin{eqnarray}
F_{2Anc} &=& 1-4p_x-4p_y-12p_z-4p_x^2-12p_xp_y \nonumber\\
	    &+& 36p_xp_z-6p_y^2+36p_yp_z+132p_z^2.
\end{eqnarray}
Note that this fidelity is higher than the fidelity when using zero or one ancilla 
only for small $p_z$. In general, I find that additional 
ancilla-based parity checks improve the fidelity with respect to $\sigma_x$ and $\sigma_y$ 
errors but reduce the fidelity with respect to $\sigma_z$ errors as shown in Fig.~\ref{fids1}. 
This is likely due to the fact that the verification is a parity check and does not check on 
the phase between the qubits.

Having constructed the noisy Shor states I now use them to perform fault syndrome measurements
so as to project an initial state of seven qubits into an encoded logical 
zero of the Steane code. I choose to utilize Shor states with the two 
verification steps used above, the first comparing qubits 1 and 4, and the second comparing 
qubits 1 and 2. A full comparison of the final encoded zero fidelity dependent 
on the choice of Shor state verification steps will be presented elsewhere. 

\section{Fault Tolerant Encoding}

To implement error correction on qubits encoded in a Steane code requires 6 
syndrome measurements. The first three of the syndrome measurements are used to check for bit-flip 
errors, using the arrangement of CNOT gates shown in Fig.~\ref{ShorState}. Then, a Hadamard 
gate is applied to each qubit, rotating phase-flips to bit-flips, and the final three syndrome 
measurements are done in the same way to check for phase-flip errors. This is followed by a final 
set of Hadamards to return the data qubits to their original basis. For a fault tolerant 
implementation of QEC each syndrome measurement uses four ancilla qubits in a Shor state 
such that one CNOT only is applied between the appropriate data qubit and a syndrome qubit.
In addition, each syndrome measurement should be done twice to account for possible errors 
in the syndrome measurement itself \cite{Preskill}.

To fault tolerantly encode a logical zero, we start with 7 qubits all in state $|0\rangle$ assumed 
to be initialized perfectly. This choice absolves us from performing the first 
set (bit-flip) of syndrome measurements as they will result in no evolution of the data 
qubits. Instead we proceed to measuring the phase flip syndrome which we would like to do with the 
least number of gates. We can eliminate all of the necessary Hadamard gates by utilitzing 
the equality pictured in the box of Fig.~\ref{ShorState}. In words, this equality states that 
applying Hadamard gates on the control and target qubits before and after application of a CNOT gate 
is equivalent to simply reversing the roles of the qubits in the CNOT gate \cite{CNOTrev}. Thus, 
we can apply the phase-flip syndrome measurement CNOT gates with the data qubits as the target and 
the ancilla qubits as the control as long as the proper Hadamard gates are applied. However, 
Hadamard gates are already necessary on the data qubits before and after 
the phase-flip syndrome measurement. Thus, reversing the CNOT gates eliminates 
these Hadamards. In addition, reversing the CNOT gates removes the need to apply the final Hadamard 
gates in the Shor state construction leaving only the need to perform Hadamards on the Shor state qubits 
after the CNOT gates. We can obviate even this need by simply measuring the Shor state qubits in the $x$-basis 
rather than the $z$-basis and in this way eliminate all Hadamard gates. 

In the actual performance of the first encoding method we join a (necessarily) noisy Shor state to the 
7 qubits all in the state $|0\rangle$ and apply the required CNOT gates as outlined above (with the ancilla 
qubits as the control). As with the construction of the Shor 
states themselves, these CNOT gates are performed in a non-equiprobable error environment such that the 
actual CNOT evolution follows Eq.~\ref{cnot}. The four syndrome qubits are measured in the $x$-basis 
and the syndrome measurement result is determined from the parity of these measured qubits. The entire 
syndrome measurement is then repeated such as to get the same result twice in a row. The repetition
is necessary to protect against the possibility of errors during the syndrome measurement itself. Since 
it is the parity of the four measured qubits that is important, there are different possible success 
outcomes each of which would lead to a slightly different final density matrix. In this paper I choose 
to analyze the scenario where all four qubits are measured as zero. 


Assuming perfect initialization of the 7 data qubits, utilizing the noisy Shor states discussed above, 
implementing the necessary (error prone) CNOT gates, selecting the case where all syndrome 
measurements give all zeros, and repeating each syndrome measurement twice, provides a full simulation 
of fault tolerant encoding of a logical zero for the $[7,1,3]$ QEC code and constructs an imperfect 
logical zero state $\rho_{7L}$. To quantify the accuracy of the encoding process we look at two distinct 
fidelity measures. The first compares the seven qubit state $\rho_{7L}$ to the ideal seven qubit logical 
zero state $|0_L\rangle$, $F_{|0_L\rangle} = \langle 0_L|\rho_{7L}|0_L\rangle$. This quantifies how
well the encoding process was implemented. For the non-equiprobable error model we find (to second order):
\begin{eqnarray}
F_{|0_L\rangle}&=& 1-48p_x-12p_y-12p_z+1236p_x^2+504p_xp_y \nonumber\\
		 &+& 528p_xp_z-596p_y^2-1795p_yp_z-1092p_z^2.
\end{eqnarray}
Here, the fidelity is most sensitive to $\sigma_x$ errors than other types of errors. 

While the fidelity of the entire seven qubit system is an appropriate accuracy measure for the 
implementation of a process, it may not be an appropriate measure for the 
accuracy with which the single qubit of quantum information is stored. Errors accounted for 
in the seven qubit fidelity may reside in degrees of freedom of the system that do not affect 
the actual stored quantum information.
To check the fidelity of the stored quantum information one may, for example, perform 
destructive measurement on all of the qubits followed by classical error correction and the 
determination of the parity of the obtained codeword. 
The probability of correctly reading out zero (in our case) would then serve as a measure of 
accuracy. Or, one may perform a parity measure of the seven qubit system using an ancilla qubit.
These accuracy measures will be explored elsewhere. Here, I apply a perfect decoding sequence 
to the seven qubit system and partial trace over qubits 2 through 7. The state of the 
remaining qubit $\rho_1$ is then compared to the single qubit state $|0\rangle$, via the fidelity 
$F_{|0\rangle} = \langle 0|\rho_{1}|0\rangle$. For fault tolerant encoding this fidelity (up to 
second order) is given by:
\begin{equation}
F_{|0\rangle} = 1-16p_x-4p_y+226p_x^2+102p_xp_y-188p_y^2.
\end{equation}
Again we see that the fidelity is most sensitive to $\sigma_x$ errors.

\section{Gate Sequence Encoding}

The second logical encoding method is implemented via the gate sequence shown in 
Fig.~\ref{ShorState}. As above, we assume perfect initialization of the 7 qubits each in the state 
$|0\rangle$ and an environment that causes non-equiprobable errors, $\sigma_x$, $\sigma_y$, and $\sigma_z$ 
on each of the qubits involved in a gate. Evolution of each of the 11 CNOT gates follows 
Eq.~\ref{cnot} and the evolution of a Hadamard gate on a state $\rho$ is given by:
\begin{equation}
\sum_{a}^{0,x,y,z}p_a\sigma_a^j H \rho H^{\dag}\sigma_a^j.
\end{equation}
When the state to be encoded is $|0\rangle$, as explored here, we can skip 
the first two CNOT gates (put in brackets in the figure) as they will have no effect on the 
target qubits. The entire encoding thus requires 3 Hadamard and 9 CNOT gates. The fidelity 
of the seven qubit logical zero state after application of this gate sequence (to second order) is:
\begin{eqnarray}
F_{|0_L\rangle}&=& 1-18p_x-21p_y-21p_z+166p_x^2+360p_xp_y \nonumber\\
		 &+& 360p_xp_z+211p_y^2+433p_yp_z+248p_z^2.
\end{eqnarray}
This fidelity is better than that achieved using the first encoding method when $\sigma_x$ 
errors are dominant, despite the fact that this second method is not fault tolerant.


The fidelity of the single qubit of stored quantum information for the encoding gate sequence is 
given by:
\begin{equation}
F_{|0\rangle} = 1-8p_x-8p_y+56p_x^2+112p_xp_y+56p_y^2.
\end{equation}
Again, with respect to this measure the gate sequence encoding is more accurate when 
$\sigma_x$ errors are dominant. 

\section{Applying Perfect Error Correction}

Applying perfect error correction allows us to test the `correctability' of the types of 
errors that occur during the encoding. If even perfect error correction cannot (to first order) 
correct the errors that occur during encoding then the encoding method cannot be 
usable for practical implementations of quantum computation. I apply perfect error correction to 
the states constructed using each of the two methods described in the previous two sections and 
calculate both fidelity measures (of the seven qubit system and the one encoded qubit).
For the fault tolerant encoding method both of the fidelities are found to be $1 - \mathcal{O}(p_j^3)$. 
For the gate sequence method, both fidelities are found to be $1-9p_x^2$.
Both encoding methods thus produce usable logical zeros for quantum computation
and the advantage of utilizing fault tolerant techniques is demonstrated by the suppression of the second
order error term. It should be stressed, however, that in realistic systems the error correction itself 
will be imperfect and thus the error correction may not correct all errors to first order. 
This will be explored in future work.

\section{Conclusion}

In conclusion, I have simulated two methods of encoding a logical zero for the $[7,1,3]$ CSS
quantum error correction code, or Steane code, in a non-equiprobable error environment. 
For both methods, each of the seven qubits are initially in the state $|0\rangle$. To implement
the first, fault tolerant, method requires the use of Shor states for syndrome measurement. 
Shor state construction involves (error prone) CNOT gates followed by verification of the 
state via parity measurements between random pairs of data qubits. While fault tolerance 
demands these verifications I have found that the fidelity of the constructed Shor state 
may decrease upon applying verification in cases where $\sigma_z$ errors are dominant. The 
noisy Shor states are then used for error syndrome measurement in the logical zero encoding.  
The second, non-fault tolerant, encoding method is to implement an error encoding gate sequence. 
Which method provides more accurate encoded logical zero states depends on the values of the 
$p_i$ characterizing the non-equiprobable error model. Finally, I apply perfect error correction
to the states constructed by the two methods and find that, to second order, all of the errors 
are corrected up to second order. This implies that either constrcution method can be used for
fault tolerant quantum computation.

I would like to thank G. Gilbert for insightful comments on this work and S. Buchbinder
and C. Huang for programming help. This research is 
supported under MITRE Innovation Program Grant 07MSR205.

\end{document}